\documentclass[aps,prd,superscriptaddress,showpacs,preprintnumbers,10pt,floatfix]{revtex4}
\usepackage{graphicx,color}
\usepackage{amssymb}

\newcommand{\be}{\begin{equation}}
\newcommand{\ee}{\end{equation}}
\newcommand{\bea}{\begin{eqnarray}}
\newcommand{\eea}{\end{eqnarray}}
\newcommand{\bq}{\begin{eqnarray}}
\newcommand{\eq}{\end{eqnarray}}

\newcommand{\bfk}{\mbox{\boldmath $k$}}
\newcommand{\bfkappa}{\mbox{\boldmath $\kappa$}}

\def\kt{k_\perp}
\newcommand{\bfp}{\mbox{\boldmath $p$}}

\newcommand{\bfh}{\mbox{\boldmath $h$}}

\newcommand{\bfP}{\mbox{\boldmath $P$}}

\def\pp{p_\perp}

\newcommand{\qup}{q^\uparrow}

\newcommand{\cosphi}{$\langle \cos \phi \rangle \, $}
\newcommand{\costwophi}{$\langle \cos 2 \phi \rangle \, $}

\newcommand{\mathtab}{\,\,\,\,\,\,\,\,\,\,\,}

\def\xb{x_{_{\!B}}}
\def\avk{\langle k_\perp ^2\rangle}
\def\avp{\langle p_\perp ^2\rangle}
\def\avPT{\langle P_T^2\rangle}

\def\C{_{_C}}
\def\BM{_{_{\rm B\!M}}}

\def\lsim{\mathrel{\rlap{\lower4pt\hbox{\hskip1pt$\sim$}}\raise1pt\hbox{$<$}}}
\def\gsim{\mathrel{\rlap{\lower4pt\hbox{\hskip1pt$\sim$}}\raise1pt\hbox{$>$}}}
\def\nostrocostruttino#1\over#2{\mathrel{\mathop{\kern 0pt \rlap
{\hbox{$#1$}}} \hbox{\kern-.135em $#2$}}}

\def\Vec#1{\hbox{\boldmath$#1$\unboldmath}}

\def\D{{\mathrm d}}
\def\E{{\mathrm e}}

\def\gev2{\,\text{GeV}^2}

\begin{document}
%%%%%%%%%%%%%%%%%%%%%%%%%%%%%%%%%%%%%%%%%%%%%%%%%%%%%%%%%%%%%%%%%%%%%%%%%%%%%%

%\preprint{}
\title{A phenomenological analysis of azimuthal asymmetries \\
in unpolarized semi-inclusive deep inelastic scattering}

\author{V. Barone}
\affiliation{Di.S.I.T., Universit\`a del Piemonte Orientale, \\
and INFN, Gruppo Collegato di Alessandria, I-15121 Alessandria, Italy}
\author{M. Boglione}
\affiliation{Dipartimento di Fisica Teorica, Universit\`a di Torino,\\
and INFN, Sezione di Torino, I-10125 Torino, Italy}
\author{J. O. Gonzalez Hernandez}
\affiliation{Dipartimento di Fisica Teorica, Universit\`a di Torino,\\
and INFN, Sezione di Torino, I-10125 Torino, Italy}
\author{S. Melis}
\affiliation{Dipartimento di Fisica Teorica, Universit\`a di Torino, I-10125 Torino, Italy}
%\date{}

\begin{abstract}

We present a phenomenological  analysis of the $\cos \phi$ and $\cos 2 \phi$ asymmetries 
in unpolarized semi-inclusive deep inelastic scattering, based on the recent multidimensional 
data released by the COMPASS and HERMES Collaborations.
In the TMD framework, valid at relatively low transverse momenta, 
these asymmetries arise from intrinsic 
transverse momentum and transverse spin effects, and from their correlations.
The role of the Cahn and Boer--Mulders effects in both azimuthal moments is explored up to order $1/Q$. 
As the kinematics of the present experiments is dominated by the low-$Q^2$ region, higher-twist contributions 
turn out to be important, affecting the results of our fits. 

\end{abstract}


\maketitle

\section{Introduction} 

Since the early QCD investigations of hadronic hard processes, it has been 
recognized that azimuthal asymmetries in unpolarized reactions, such as Drell-Yan production 
and semi-inclusive deep inelastic scattering (SIDIS), represent an important window 
on the perturbative and non-perturbative aspects of strong interactions \cite{Georgi:1977tv,Mendez:1978zx}.  
Focusing on SIDIS processes,  while at large momentum transfer $Q$ and large $P_T$ (the transverse momentum 
of the produced hadron), the azimuthal asymmetries are perturbatively 
generated by gluon radiation, at small $P_T$, $P_T \ll Q$, they 
can arise from the intrinsic motion of quarks \cite{Cahn:1978se,Cahn:1989yf,Konig:1982uk,Chay:1991nh} 
(for a review, see Ref.~\cite{Barone:2010zz}). 
In the latter regime, the so-called transverse-momentum-dependent (TMD) factorization 
applies, and the SIDIS structure functions (in the current fragmentation region) 
can be written as transverse-momentum convolutions
of TMD distribution and fragmentation functions \cite{Ji:2004wu}. 

Two azimuthal modulations appear in the unpolarized SIDIS cross 
section, of the type $\cos \phi$ and $\cos 2 \phi$, where $\phi$ is the azimuthal 
angle of the produced hadron (measured from the scattering plane). 
These asymmetries, which have been experimentally investigated for the first time 
in the large $Q^2$ region by the EMC and ZEUS experiments
\cite{Aubert:1983cz,Arneodo:1986cf,Breitweg:2000qh},  
have recently attracted a large experimental and theoretical attention as 
a potential source of information,  in the small $P_T$ region,
on the so-called 
Boer--Mulders distribution function, $h_1^{\perp}$, which measures 
the transverse polarization asymmetry of quarks inside an unpolarized 
nucleon~\cite{Boer:1997nt}.  
  
A few years ago the azimuthal asymmetries in unpolarized SIDIS 
have been measured by the COMPASS and HERMES Collaborations 
for positive and negative hadrons, and presented as one-dimensional projections, 
with all variables $(x_B, z_h, Q^2, P_T)$ but one 
integrated over~\cite{Kafer:2008ud,Bressan:2009eu,Giordano:2009hi}.

The one-dimensional data on the $\cos 2 \phi$ asymmetry were 
analyzed in Ref.~\cite{Barone:2009hw}, where it was shown  
that the larger asymmetry for $\pi^-(h^-)$ production, 
compared to $\pi^+(h^+)$, was an indication of the existence of a non-zero Boer--Mulders effect, 
in agreement with the earlier predictions of Ref.~\cite{Barone:2008tn}. 
It was also pointed out that measurements at different values of $Q^2$ were essential,
in order to disentangle higher-twist contributions from the twist-two Boer-Mulder term.

The HERMES and COMPASS 
Collaborations have recently provided multidimensional data 
in bins of $x_B, z_h, Q^2$ and $P_T$ 
for the multiplicities~\cite{Airapetian:2012ki,Adolph:2013stb} and for the azimuthal 
asymmetries~\cite{Airapetian:2012yg,Adolph:2014pwc}. 
In principle, multidimensional data should offer detailed information, 
essential to unravel the kinematical behaviour of these asymmetries: 
for instance, the $Q^2$ dependence is crucial to disentangle higher-twist effects.   
In this paper, we present a study of the SIDIS azimuthal moments 
\cosphi and \costwophi in order to understand the role of the Cahn effect and to 
extract the Boer--Mulders function. We will see that, due to the present kinematics 
which is still dominated by the low-$Q^2$ region, the higher-twist contributions 
are important and strongly affect the results of our fits.

%%%%%%%%%%%%%%%%%%%%%%%%%%%%%%%%%%%%%%%%%%%%%%%%%%%%%%%%%%%%%%%%%%%%%%
\section{Formalism}
\label{formalism}
%%%%%%%%%%%%%%%%%%%%%%%%%%%%%%%%%%%%%%%%%%%%%%%%%%%%%%%%%%%%%%%%%%%%%%

The process we are interested in is unpolarized SIDIS:
\begin{equation}
l (\ell) \, + \, N (P) \, \rightarrow \, l' (\ell')
\, + \, h (P_h) \, + \, X (P_X)\,.
\label{sidis}
\end{equation}
The cross section of this process is expressed in terms 
of the invariants
\begin{equation}
x_B = \frac{Q^2}{2 \, P \cdot q}, \;\;\;
y =  \frac{P \cdot q}{P \cdot \ell} ,
\;\;\;
z_h = \frac{P \cdot P_h}{P \cdot q}\,,
\end{equation}
where $ q = \ell - \ell'$ and $Q^2 \equiv - q^2$. 

%%%%%%%%%%%%%%%%%%%%%%%%%%%%%%%%%%%%%%%%%%%%%%%%%%%%%%%%%%%%%%%%%%%%%%%%%%%%%%%%%%%%%%%%%%%%%%%%%%%%%%
\begin{figure}[t]
\centering
\includegraphics[width=0.5\textwidth]{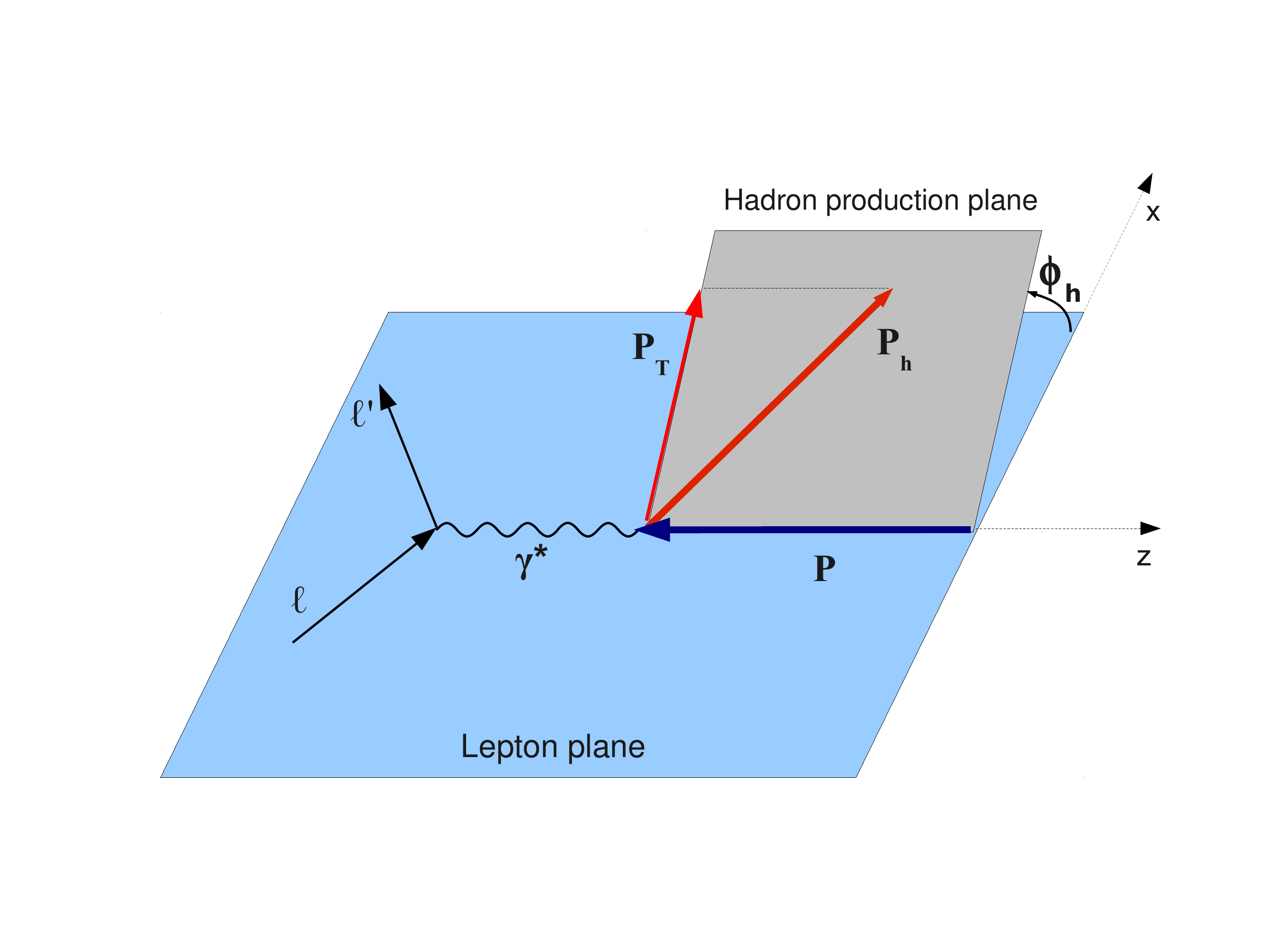}
\caption{%\label{sidisfig}
Kinematical configuration and conventions for SIDIS processes.
The initial and final lepton momenta define the $\hat{x}$-$\hat{z}$ plane.}\label{sidisfig}
\end{figure}
%%%%%%%%%%%%%%%%%%%%%%%%%%%%%%%%%%%%%%%%%%%%%%%%%%%%%%%%%%%%%%%%%%%%%%%%%%%%%%%%%%%%%%%%%%%%%%%%%%%%%%

The reference frame we adopt is the $\gamma^*$-$N$ 
center-of-mass frame, with the virtual
photon moving in the positive $z$ direction 
(Fig.~\ref{sidisfig}). 
We denote by $\bfP_T$ the transverse momentum of the produced hadron. 
The azimuthal angle of this hadron referred to the lepton scattering plane 
will be called $\phi$. 

The unpolarized SIDIS differential cross section is 
\begin{eqnarray}
\frac{\D\sigma}{\D x_B \D y \D z_h \D P_T^2 \D \phi} &=&
\frac{\pi \alpha^2}{Q^2 x y}\left \{(1+(1-y)^2) F_{UU}+2(2-y)\sqrt{1-y}F_{UU}^{\cos\phi}\cos\phi
\right. \nonumber \\
& & + \left. 2(1-y)F_{UU}^{\cos(2\phi)}\cos(2\phi)\right \}, 
\label{sidis-crosssection}
\end{eqnarray}
where the structure functions $F_{UU}, F_{UU}^{\cos \phi}, F_{UU}^{\cos 2 \phi}$ 
depend on $x_B, z_h, Q^2, P_T^2$. $F_{UU}$ 
is the structure function which survives upon integration  
over $\phi$, while $F_{UU}^{\cos \phi}$ and $F_{UU}^{\cos 2 \phi}$ 
are associated to the $\cos \phi$ and $\cos 2 \phi$ modulations, 
respectively. 

If $k$ is the momentum of the quark inside the proton, 
and $\bfk_{\perp}$ its transverse component with respect to 
the $\gamma^* N$ axis, in the kinematical region where 
$P_T \sim k_{\perp} \ll Q$, the transverse-momentum-dependent 
(TMD) factorization is known to hold at leading twist. 
The structure functions can be expressed in terms of TMD distribution 
and fragmentation functions, which depend on the light-cone 
momentum fractions
\be
x = \frac{k^+}{P^+}\,, \;\;\;
z = \frac{P_h^-}{\kappa^-},  
\label{lightcone}
\ee
where $\kappa$ is the momentum of the fragmenting quark. 
Although the TMD factorization has not been rigorously proven at order $1/Q$ 
(that is, at twist three), in the following we will assume it to hold, 
as usually done in most phenomenological analyses.  

Introducing the 
transverse momentum $\bfp_{\perp}$ of the final hadron 
with respect to the direction of the fragmenting quark,  
up to order $k_{\perp}/Q$ one has $\bfp_{\perp} 
= - z \bfkappa_{\perp}$ 
and the momentum conservation reads
\be
\Vec P_T = z \bfk_{\perp} + \bfp_{\perp}.   
\label{transv_conserv} 
\ee
At the same order we can identify the light-cone momentum fractions 
with the invariants $x_B$ and $z_h$, 
\be
x = x_B , \;\;\;\;
z = z_h. 
\label{xz}
\ee
In the TMD factorization scheme the structure function $F_{UU}$ is given by 
\be
F_{UU}=\sum_q e_q^2 x \int \D^2 \Vec k_{\perp} \, f_q(x, \kt) D_q(z,p_\perp), 
\label{FUU}
\ee
where $f_q(x, k_{\perp})$ and $D_q (z, p_{\perp})$ are the unpolarized TMD 
distribution and fragmentation function, respectively, for the flavor $q$ (the sum 
is intended to be both over quarks and antiquarks). Note that in Eq.~(\ref{FUU}) 
the transverse momentum conservation has been applied, so that 
$p_{\perp} = \vert \Vec P_T - z \bfk_{\perp} \vert$. The $Q^2$ dependence of all 
functions is omitted for simplicity. 

The structure function associated with the $\cos \phi$ modulation 
turns out to be an order $1/Q$, i.e. a twist-3, quantity. 
Neglecting the dynamical twist-3 contributions (the so-called 
``tilde'' functions,  which arise from quark-gluon correlations), 
$F_{UU}^{\cos \phi}$ can be written as the sum of two terms  
\be
F_{UU}^{\cos \phi} = \left. F_{UU}^{\cos \phi} \right \vert_{\rm Cahn} + 
\left. F_{UU}^{\cos \phi} \right \vert_{\rm BM}, 
\ee
with  ($\bfh \equiv \bfP_T/\vert \bfP_T \vert$)
\bq
& & \left. F_{UU}^{\cos\phi} \right \vert_{\rm Cahn} = 
- 2 \sum_q e_q^2 x \int \D^2\bfk_{\perp}\,\frac{(\bfk_{\perp}\cdot \bfh)}{Q}f_q(x,\kt) D_q(z,p_\perp), 
\label{cosphiCahn} \\
& &  \left. F_{UU}^{\cos\phi}\right \vert_{\rm BM}
=\sum_q e_q^2 x \int \D^2\bfk_{\perp}\,\frac{\kt}{Q} 
\frac{P_T-z(\bfk_{\perp}\cdot \bfh)}{k_{\perp}} \Delta f_{\qup/p}(x, \kt) \Delta  
D_{h/q^\uparrow}(z,p_\perp).  
\label{cosphiBM}
\eq
Eq.~(\ref{cosphiCahn}) is the Cahn term, which accounts for the 
non-collinear kinematics of quarks in the elementary subprocess $\ell q \to \ell' q'$. 
Eq.~(\ref{cosphiBM}) is the Boer--Mulders contribution,  arising from the  
correlation between the transverse spin and the transverse momentum  
of quarks inside the unpolarized proton. In this term the Boer--Mulders 
distribution function $\Delta f_{\qup}/p$ couples to the Collins fragmentation 
function $\Delta D_{h/\qup}$. The relations between these functions, as defined 
in the present paper, and the 
corresponding quantities in the Amsterdam notation is 
\bq
& & \Delta f_{\qup/p}(x, \kt) 
= - \frac {\kt}{M_p} \, h_{1}^\perp (x, \kt), 
 \label{b-m} \\
& & \Delta  D_{h/q^\uparrow}(z,p_\perp) =
\frac{2\, p_{\perp}}{z M_h} H_{1}^{\perp}(z,p_{\perp}),
\label{D-prop}
\eq
where $M_p$ and $M_h$ are the masses of the proton and of the final hadron, respectively. 
The Boer--Mulders effect is also responsible of the $\cos 2 \phi$ 
modulation of the cross section, giving a leading--twist contribution
(that is, unsuppressed 
in $Q$). It has the form
\bq
\left. F_{UU}^{\cos2\phi} \right \vert_{\rm BM} & = & - \sum_q e_q^2 x \int \D^2\bfk_{\perp}\,
\frac{P_T (\bfk_{\perp}\cdot \bfh)+ z_h \left[ \kt^2 -2 (\bfk_{\perp}\cdot \bfh)^2\right]}{2 
\, k_{\perp} p_{\perp}} \nonumber \\
& & \times \Delta f_{\qup/p}(x, \kt) \Delta D_{h/q^\uparrow}(z,p_\perp). 
\label{cos2phiBM}
\eq
Concerning higher-twists, the $\cos 2 \phi$ structure function has no $1/Q$ component, but 
receives various $1/Q^2$ contributions. Only one of these, of the Cahn type, is known  
and reads
\be
\left. F_{UU}^{\cos2\phi} \right \vert_{\rm Cahn}
= 2\sum_q e_q^2 x \int \D^2\bfk_{\perp}\, 
\frac{2(\bfk_{\perp}\cdot \bfh)^2-\kt^2}{Q^2}f_q(x,\kt) D_q(z,p_\perp). 
\label{cos2phiCahn}
\ee
Notice however that at order $1/Q^2$ many simplifying kinematical 
relations do not hold any longer (for instance, $x$ and $z$ do not 
coincide with $x_B$ and $z_h$), and there appear target mass corrections. 
Thus, Eq.~(\ref{cos2phiCahn}) must be intended only as an approximation 
to the full twist-4 contribution to $F_{UU}^{\cos 2 \phi}$.

For the TMD functions we will use a factorized form, with the 
transverse-momentum dependence modeled by Gaussians. This Ansatz 
is supported by phenomenological analyses (see for instance Ref.~\cite{Schweitzer:2010tt})  
and lattice simulations \cite{Musch:2007ya}.
Thus, the unpolarized distribution and fragmentation functions are parametrized as
 \bq
&&f_{q/p} (x,\kt)= f_{q/p} (x)\,\frac{\E^{-\kt^2/\avk}}{\pi\avk}, 
\label{unp-dist}\\
&&D_{h/q}(z,\pp)=D_{h/q}(z)\,\frac{\E^{-\pp^2/\avp}}{\pi\avp}. 
\label{unp-frag}
\eq
The integrated functions, $f_{q/p}(x)$ and $D_{h/q}(z)$, will be taken from the 
available fits of the world data (in particular, we will use the CTEQ6L set 
for the distribution functions \cite{Pumplin:2002vw} and the DSS set for the fragmentation 
functions~\cite{deFlorian:2007aj}). The widths of the Gaussians might  
depend on $x$ or $z$, and be different for different distributions: we will 
discuss these possibilities later on.  

For the Boer--Mulders function we use the following parametrization
\be
\Delta f_{q^\uparrow/p}(x,\kt)= \Delta f_{q^\uparrow/p} (x)\,  
 \sqrt{2e}\,\frac{k_\perp}{M \BM}\,\E^{-{k_\perp^2}/{M \BM}^2}\,  \frac{\E^{-\kt^2/\avk}}{\pi\avk}, 
\label{BMfunct}
\ee
with 
\be
\Delta f_{q^\uparrow/p} (x) = 
N_q \, 
\frac{(\alpha_q+\beta_q)^{(\alpha_q+\beta_q)}}{\alpha_q^{\alpha_q} \beta_q^{\beta_q}} \, 
x^{\alpha_q}(1-x)^{\beta_q} \,
\, f_{q/p} (x),
\label{BMfunct2}
\ee
where $\alpha_q$, $\beta_q$ and $M \BM$ are free parameters to be determined by the fit.
The distribution so constructed is such that the positivity bound 
$\vert \Delta f_{q^\uparrow/p} (x, k_{\perp}) \vert \leq 2 \, f_{q/p} (x, k_{\perp})$ 
is automatically satisfied. Multiplying the two Gaussians in 
Eq.~(\ref{BMfunct}), the Boer--Mulders function can be rewritten as 
\be
 \Delta  f_{q^\uparrow/p}(x, k_{\perp}) = 
\Delta f_{q^\uparrow/p} (x) \, 
\sqrt{2e}\,\frac{\kt}{M \BM} \; \frac{e^{-\kt^2/\avk \BM}}{\pi\avk},  
\label{BM-dist}
\ee
with 
\be
\avk \BM = \frac{\avk \, M^2 \BM}{\avk + M^2 \BM}. 
\ee
For the Collins function we have a similar parametrization, namely 
\be
\Delta  D_{h/q^\uparrow}(z,\pp) = \Delta  D_{h/q^\uparrow}(z) \, 
\sqrt{2e}\,\frac{\pp}{M_C} \; e^{-\pp^2/M_C^2}\,
\frac{e^{-\pp^2/\avp}}{\pi\avp}, 
\label{Collfunct}
\ee
with 
\be
\Delta D_{h/q^\uparrow} (z) = N^C_q 
\,\frac{(\gamma+\delta)^{(\gamma+\delta)}}{\gamma^{\gamma} \delta^{\delta}}
\, z^{\gamma}(1-z)^{\delta} 
\, D_{h/q}(z),  
\ee
where $\gamma$, $\delta$ and $M_C$ are free parameters.
Combining the two Gaussians in (\ref{Collfunct}), we get 
\be
\Delta  D_{h/q^\uparrow}(z,\pp) = 
\Delta  D_{h/q^\uparrow}(z) \, 
\sqrt{2e}\,\frac{\pp}{M_C} \;
\frac{e^{-\pp^2/\avp \C}}{\pi\avp}\,,\label{Coll-frag}
\ee
having defined
\be
\avp \C=\frac{\avp \, M_C^2}{\avp +M_C^2}. 
 \label{Coll-frag2}
\ee
The Gaussian parametrization has the advantage that the integrals 
over the transverse momenta can be performed analytically. Thus,  
inserting the above distribution and fragmentation functions into 
the expressions for the SIDIS structure functions, we get
\bq
& & F_{UU}  =  \sum_{q} \, e_q^2 \,x_B \, f_{q/p}(\xb)\,D_{h/q}(z_h)
\frac{e^{-P_T^2/\avPT}}{\pi\avPT},  \label{G-FUU}\\
& & F_{UU}^{\cos\phi}|_{\rm Cahn}  =  -2\,\frac{P_T}{Q}\, \sum_{q} \, e_q^2 \, x_B \, 
f_{q/p}(\xb) \, D_{h/q}(z_h) \, \frac{z_h \avk}{\avPT}\,\frac{e^{-P_T^2/\avPT}}{\pi\avPT}, 
\label{g-FUUcosphi_cahn}
\\
& & \left. F_{UU}^{\cos\phi} \right \vert_{\rm BM}  = 2 e\,\frac{P_T}{Q}\,\sum_{q} \, e_q^2 \, x_B \, 
\frac{ \Delta f_{q^\uparrow/p}(\xb)}{M \BM}\, \frac{\Delta  D_{h/q^\uparrow}(z_h)}
{M_C} \, \frac{e^{-P_T^2/\avPT \BM}}{\pi\avPT ^4 \BM} \,\\
&& \hspace{2cm}  \times
\frac{\avk^2 \BM \avp ^2 \C}{\avk \avp} \Big[z_h^2 \avk \BM \Big(P_T^2 -
\avPT \BM \Big) + \avp \C \avPT \BM \Big], 
\label{g-FUUcosphi} 
%\\
\eq
\bq
& & \left. F_{UU}^{\cos2\phi} \right \vert_{\rm Cahn}  =  2\,\frac{P_T^2}{Q^2}\, \sum_{q} \, e_q^2 \, x_B \, 
f_{q/p}(\xb) \, D_{h/q}(z_h) \, \frac{e^{-P_T^2/\avPT}}{\pi\avPT}\, \frac{z_h^2 \avk^2}{\avPT^2}, 
\label{g-FUUcos2phi_cahn} 
\\
& & \left. F_{UU}^{\cos2\phi} \right \vert_{\rm BM}  =  -e P_T^2 \, \sum_{q} \, e_q^2 \, x_B \, 
\frac{ \Delta f_{q^\uparrow/p}(\xb)\,\Delta  D_{h/q^\uparrow}(z_h)}{M \BM M_C} \,
\nonumber\\
&&  \hspace{2cm} \times \frac{e^{-P_T^2/\avPT \BM}}{\pi\avPT ^3 \BM}
\, \frac{z_h\,\avk ^2 \BM \avp ^2 \C}{\avk \avp }, 
\label{g-FUUcos2phi}
\eq
where
\be
\avPT = \avp + z_h^2 \, \avk,  
\label{rel_transv} 
\ee
and 
\be
 \avPT_{BM} = \avp \C + z_h^2 \, \avk_{BM} \>. 
\label{rel_transv_BM}
\ee
The quantities actually measured in unpolarized SIDIS experiments are 
the multiplicities and the azimuthal asymmetries. 
The differential hadron multiplicity is defined as 
\be
\frac{\D^2 n^h}{\D z_h \, \D P_T^2} = 
\left ( \frac{\D^2 \sigma^{\rm DIS}}{\D \xb \, \D y} \right )^{-1} \, 
\frac{\D^4 \sigma}{\D \xb \, \D y \, \D z_h \, \D P_T^2} ,
\label{mult-c}
\ee
The deep inelastic scattering (DIS) cross section has the usual leading-order 
collinear expression  
\be
\frac{\D^2 \sigma^{\rm DIS}}{\D\xb \, \D y} = 
\frac {2 \, \pi \, \alpha^2}{\xb^2 s} \, \frac{\left[ 1 + (1-y)^2 \right]}{y^2}
\sum_{q} e_q^2 x_B f_{q/p} (\xb) \label{xs-DIS}. 
\ee
Inserting  the SIDIS cross section 
(\ref{sidis-crosssection}) integrated over $\phi$,  and Eqs.~(\ref{G-FUU}) and (\ref{xs-DIS}) 
into Eq.~(\ref{mult-c}), we find for the multiplicities 
\be
\frac{\D^2 n^h}{\D z_h \, \D P_T^2} = 
\frac{\pi \> \sum_{q} \, e_q^2 \,x_B f_{q/p}(\xb)\,D_{h/q}(z_h)}
{\sum_{q} e_q^2 x_B f_{q/p} (\xb)} 
\frac{e^{-P_T^2/\avPT}}{\pi\avPT}
\>, \label{mult-gaus}
\ee
The $\cos \phi$ and $\cos 2 \phi$ asymmetries are defined as 
\begin{eqnarray}
& & A^{\cos \phi} \equiv 2\langle \cos \phi\rangle=2\frac{\int  \! \D\phi\, 
\D \sigma \cos\phi}{\int \D\phi \, \D\sigma  },  \\
& & A^{\cos 2\phi} \equiv 2\langle \cos 2\phi\rangle=2\frac{\int \! \D\phi\, 
\D\sigma \cos2\phi}{\int \D\phi\, \D\sigma  }, 
\end{eqnarray}
that is, in terms of the structure functions, 
\begin{eqnarray}
& & A^{\cos \phi}=\frac{ 2(2-y)\sqrt{1-y}}{ \left[ 1+(1-y)^2 \right]}
\frac{ F_{UU}^{\cos\phi}}{ F_{UU}},  \\
& & A^{\cos 2\phi}=\frac{ 2(1-y)}{ \left[ 1+(1-y)^2 \right]}
\frac{ F_{UU}^{\cos2\phi}}{ F_{UU}}. 
\end{eqnarray}

In the following we will fit simultaneously the multiplicities and the 
$\cos \phi$ and $\cos 2 \phi$ asymmetries. As seen from Eq.~(\ref{mult-gaus}), 
the data on multiplicities, 
which statistically dominate our dataset, constrain $\avPT$ 
only, which in the Gaussian model (and neglecting $1/Q^2$ corrections) 
is given by the combination $\avPT = z^2 \avk + \avp $, see Eq.~(\ref{rel_transv}). 
The azimuthal asymmetries, 
on the other hand, depend on $\avk$ and $\avp$ separately, offering the chance to gain information 
on the individual values of these two average intrinsic momenta.

%%%%%%%%%%%%%%%%%%%%%%%%%%%%%%%%%%%%%%%%%%%%%%%%%%%%%%%%%%%%%%%%%%%%%%

\section{Analysis of \cosphi and \costwophi} 

%%%%%%%%%%%%%%%%%%%%%%%%%%%%%%%%%%%%%%%%%%%%%%%%%%%%%%%%%%%%%%%%%%%%%%

We will use the HERMES and COMPASS
multidimensional data, which are provided  
in bins of $x_B, z_h, Q^2$ and $P_T$ 
for the multiplicities~\cite{Airapetian:2012ki,Adolph:2013stb} and for the azimuthal 
asymmetries~\cite{Airapetian:2012yg,Adolph:2014pwc}. 
The HERMES multiplicity measurements~\cite{Airapetian:2012ki} refer to identified 
hadron productions ($\pi^+$, $\pi^-$, $K^+$, $K^-$) off proton and 
deuteron targets and cover the kinematical region of $Q^2$ values between 
$1$ and $10$ GeV$^2$ and $0.023 \le \xb \le 0.6$, while the COMPASS multiplicity data~\cite{Adolph:2013stb} refer 
to unidentified charged hadron production 
($h^+$ and $h^-$) off a deuteron target ($^6$LiD), cover the region
$0.0045 \le \xb \le 0.12$ and are binned in a $Q^2$ region similar to that of the HERMES experiment.  
In Ref.~\cite{Airapetian:2012yg}, the HERMES collaboration released azimuthal asymmetries for unpolarized SIDIS 
identified hadron production off proton and deuteron targets.
Multidimensional data on the $A^{\cos \phi}$ and $A^{\cos 2\phi}$ asymmetries are provided, binning the data 
in the kinematical variables $x_B$, $y$, $z_h$ and $P_T$. 
The kinematical cuts used for their analysis are the following
\begin{equation}
 0.023<x_B<0.6\,,\,\,\,0.2<z_h<1.0\,,\,\,\,0.05 \gev2 <P_T<1.3 \gev2\,, 
\end{equation}
 with the additional constraints
\begin{equation}
 Q^2 > 1.0\,\gev2\,,\,\,\, W^2 > 10.0\,\gev2\,,\,\,\, 0.2 < y < 0.85
\end{equation}
binned according to Table~II of Ref.~\cite{Airapetian:2012yg}. 
One dimensional projections of the azimuthal moments are also provided, 
integrated on the kinematical regions presented in Table~III of Ref.~\cite{Airapetian:2012yg}.

The COMPASS collaboration azimuthal asymmetries for unpolarized SIDIS 
charged hadron-production off a deuteron target are presented in Ref.~\cite{Adolph:2014pwc}.
In their analysis, both one-dimensional and multidimensional versions of $A^{\cos \phi}$ and $A^{\cos 2\phi}$ 
were extracted, binning the data in the kinematical variables $x_B$, $z_h$ and $P_T$. 
The kinematical cuts used for their analysis are the following
\begin{equation}
 0.003<x_B<0.13\,,\,\,\,0.2<z_h<0.85\,,\,\,\,0.1 \gev2 <P_T<1.0 \gev2\,, 
\end{equation}
 with the additional constraints
\begin{equation}
 Q^2 > 1.0\,\gev2\,,\,\,\, W^2 > 25.0\,\gev2\,,\,\,\, 0.2 < y < 0.9\,.
\end{equation}

In the following, we are going to study the multidimensional  SIDIS azimuthal moments 
\cosphi and \costwophi with the aim of understanding the role of the Cahn and Boer--Mulders effects. 
We expect multidimensional data to be useful to explore 
the kinematical behavior of these asymmetries, and their explicit $Q^2$ dependence 
to be of help in exploring the interplay between 
leading-twist and higher-twist contributions.   
We perform our analysis by including contributions up to order $\mathcal{O}(1/Q)$;
the role of dynamical twist--three contributions will be briefly discussed at the end of the Section.

Up to order $\mathcal{O}(1/Q)$, \cosphi receives 
contributions from the Cahn and the Boer--Mulders effect, which appear at 
subleading twist in $F_{UU}^{\cos\phi}$, Eqs.~(\ref{cosphiCahn}) and (\ref{cosphiBM}), 
while \costwophi is proportional to the sole Boer--Mulders effect, 
which instead appears at leading twist, Eq.~(\ref{cos2phiBM}). 
Both asymmetries contain, at denominator, the contribution of $F_{UU}$, 
defined in Eq.~(\ref{FUU}).

$F_{UU}$ and the Cahn contribution to $\langle \cos \phi \rangle$ involves only the 
unpolarized TMD distribution and fragmentation functions $f_{q/p}(x, k_{\perp})$ 
and $D_{h/q} (z, p_{\perp})$. These functions have been recently extracted in Ref.~\cite{Anselmino:2013lza}, 
from a best fit of HERMES and COMPASS multidimensional multiplicity data. In this analysis, 
a Gaussian model was used for the $\kt$ and $\pp$ dependence as in Eqs.~(\ref{unp-dist}) and~(\ref{unp-frag}).

If we use the values of $\avk$ and $\avp$ obtained there
we find a very large Cahn contribution, of the order of $50\%$ 
which largely overshoots the data. This is not surprising. 
Eq.~(\ref{g-FUUcosphi_cahn}) shows in fact that $F_{UU}^{\cos \phi} \vert_{\rm Cahn}$ 
is proportional to $\avk$, which was found to be rather large, about 0.6 GeV$^2$. 
It is quite unlikely that any reasonable 
Boer--Mulders contribution could cancel this huge Cahn term so as to reproduce  
the observed \cosphi asymmetry,  which does not exceed $10$\%. 

However, in Ref.~\cite{Anselmino:2013lza} it was observed that, 
since the multiplicities are sensitive only to the combination 
$\avPT= z_h^2 \avk + \avp $, Eq.~(\ref{G-FUU}), they cannot distinguish $\avk$ from  $\avp$.
Instead, the azimuthal asymmetries 
involve $\avk$ and $\avp$ separately, and are sensitive to a $z_h$-dependent $\avp$, 
see Eqs.~(\ref{g-FUUcosphi_cahn})--(\ref{g-FUUcos2phi}). 
For example, if one takes $\avk$ as a free constant parameter, $\avk = C$,  
and allows $\avp$ to have a quadratic $z_h$ dependence of the form
\be
\avp = A + B z_h^2\, , 
\label{avp_z}
\ee
where $A$ and $B$ are two additional constant parameters, one finds 
\be
\avPT = A + (B + C) z_h^2 \,,
\label{BCA} 
\ee
which is of the same form of Eq.~(\ref{rel_transv}). A fit of the multiplicities using this functional 
form would lead to the same results, but would allow for a different interpretation of the extracted 
parameters in terms of $\avk$ and $\avp$.
Instead if, in addition, we fit also $\langle \cos \phi \rangle$ and $\langle \cos 2 \phi \rangle$ data, 
we acquire some degree of sensitivity to the individual values of A, B and C.
Here, in principle, C can be small and lead to a much smaller Cahn contribution.

%%%%%%%%%%%%%%%%%%%%%%%%%%%%%%%%%%%%%%%%%%%%%%%%%%%%%%%%%%%%%%%%%%%%%
\begin{figure}[t]
    \centering
\includegraphics[width=11.3cm]{COMPASSAZIM-2cosphi-vs-PT-main-fit}
    \caption{Best fit curves for \cosphi obtained by fitting COMPASS data
    on multiplicities, \cosphi and \costwophi. 
The Cahn effect in \costwophi has been set to zero.}
    \label{plot:cosphi:mainfit-C}
\end{figure}
%%%%%%%%%%%%%%%%%%%%%%%%%%%%%%%%%%%%%%%%%%%%%%%%%%%%%%%%%%%%%%%%%%%%%%
%%%%%%%%%%%%%%%%%%%%%%%%%%%%%%%%%%%%%%%%%%%%%%%%%%%%%%%%%%%%%%%%%%%%%%
\begin{figure}[t]
    \centering
\includegraphics[width=11.3cm]{COMPASSAZIM-2cos2phi-vs-PT-main-fit}
    \caption{Best fit curves for \costwophi obtained by fitting COMPASS data
    on multiplicities, \cosphi and \costwophi. 
The Cahn effect in \costwophi has been set to zero.}
    \label{plot:cos2phi:mainfit-C}
\end{figure}
%%%%%%%%%%%%%%%%%%%%%%%%%%%%%%%%%%%%%%%%%%%%%%%%%%%%%%%%%%%%%%%%%%%%%%

 In this paper we explore this configuration. We perform a 
 global best fit which includes 
 the multiplicities, the $\cos \phi$ asymmetry and 
 the $\cos 2 \phi$ asymmetry. Working up to order $1/Q$, these asymmetries read:
\[ %\left\{
  \begin{array}{l l}
  A^{\cos \phi} = \left. A^{\cos \phi} \right \vert_{\rm Cahn} + 
 \left. A^{\cos \phi} \right \vert_{\rm BM}
\\
\\
    A^{\cos 2 \phi} = \left. A^{\cos 2 \phi} \right \vert_{\rm BM}
  \end{array} %\right.
  \]
As both COMPASS and HERMES data on $\langle \cos \phi \rangle$ and
$\langle \cos 2 \phi \rangle$ are restricted to a 
narrow $x$ range, they do not allow us to determine the precise $x$-dependence of the Boer--Mulders function. 
Thus we take $\Delta f_{q^\uparrow/p}$ to be simply proportional 
to $f_{q/p}$, by setting $\alpha_q = \beta_q = 0$ in Eq.~(\ref{BMfunct2}).  
For the Collins function, we distinguish a favored and a disfavored component,  
and we fix their parameters to the values 
obtained in a recent fit of the Collins asymmetries 
in SIDIS and $e^+ e^-$ annihilation~\cite{Anselmino:2013vqa}: 
 \bq
 & & N_{\rm fav}^C = 0.49, \;\;\;
 N_{\rm disf}^C = -1.00, \nonumber \\
 & & \gamma= 1.06, \;\;\;
 \delta = 0.07, \nonumber \\
 & & M_C^2 = 1.50 \; {\rm GeV}^2. 
 \eq 
 To parameterize the Boer--Mulders and Collins functions
 we need to input the unpolarized $f_{q/p}(x)$ and $D_{q/p}(z)$, 
 see Eqs.~(\ref{BMfunct2}) and~(\ref{Coll-frag}). Consistently with our previous choice, 
 Eqs.~(\ref{unp-dist}) and (\ref{unp-frag}), we will use the collinear CTEQ6L distribution 
 functions~\cite{Pumplin:2002vw} 
 and DSS fragmentation functions ~\cite{deFlorian:2007aj}.
 As mentioned before, $\avk=C$ and $\avp=A + B z_h^2$, with $A$, $B$ and $C$
free parameters to be determined by the fit. 

It is known that the COMPASS multiplicities should be corrected by a large normalization factor: 
in fact, issues in that analysis were detected, which can affect the overall $x$, $y$, $z$ normalization of
multiplicities up to 40\%, as pointed out in Ref.~\cite{Stolarski:2015}. Lacking 
 for the moment the corrected data, in the 
 present paper we apply the same multiplicative normalization factor as obtained in~\cite{Anselmino:2013lza},
 \(
 N_y = 1.06 - 0.43 \,  y .
 \label{corr_y}
 \)
This correction was found to 
 improve considerably the fit of the COMPASS multiplicities. 
 
The kinematical range explored by the two experiments is further restricted in order to 
make sure that our description, based on TMD factorization, can safely be applied. 
To avoid contaminations from exclusive 
hadronic production processes and large $z$ resummation 
effects~\cite{Anderle:2012rq} we select data with $z_h < 0.65$ for COMPASS and with $z_h < 0.69$ for HERMES.
The lowest cut in $Q^2$ is chosen accordingly to the minimum  $Q^2$ in the 
CTEQ6L analysis, $Q^2 > 1.69$ GeV$^2$, which amounts to excluding the lowest 
$x$ bins.
Finally, we select $0.2 < P_T < 0.9 \; {\rm GeV} $, following Ref.~\cite{Anselmino:2013lza}.

\begin{figure}[t]
    \centering
\includegraphics[width=13.5cm]{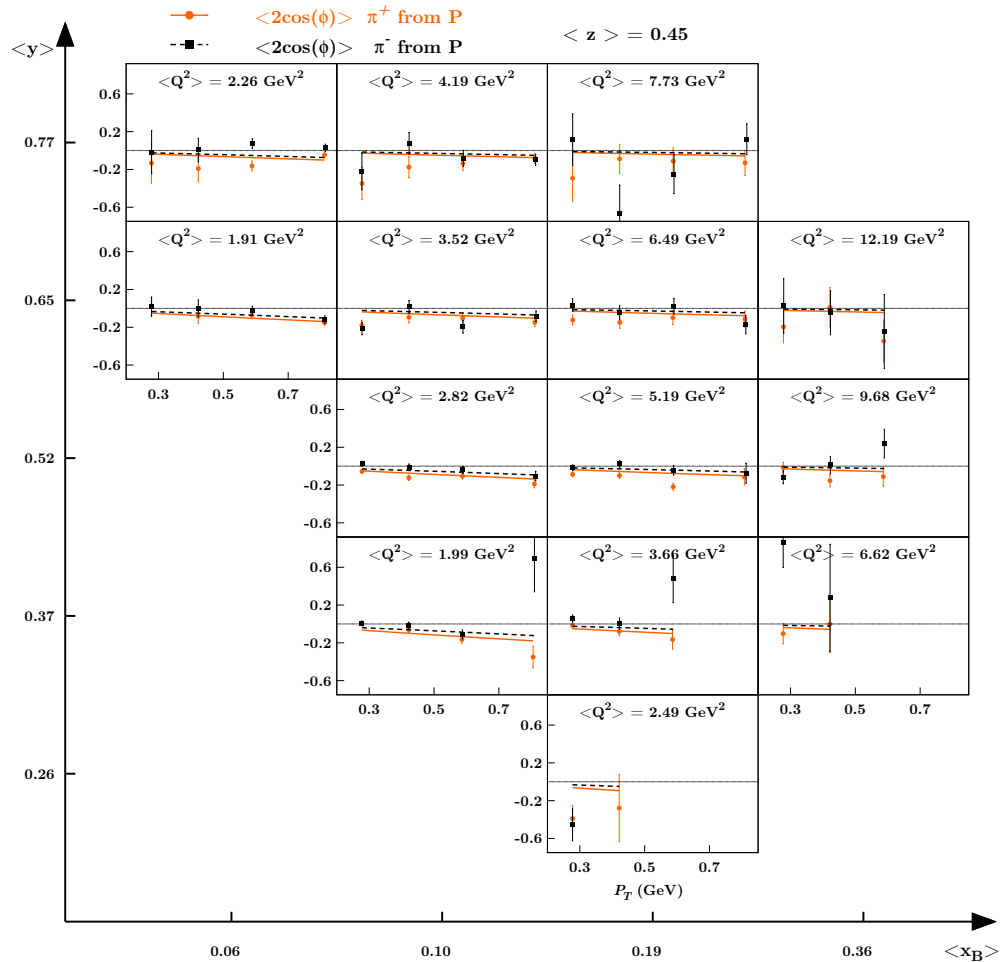}
    \caption{Best fit curves for \cosphi obtained by fitting HERMES data
    on multiplicities, \cosphi and \costwophi. 
    The Cahn effect in \costwophi has been set to zero. Here we show only one bin in $z$, as an example, with $\langle z \rangle = 0.45$}
    \label{plot:cosphi:mainfit-H}
\end{figure}

The results of the fit for the azimuthal moments are shown in Figs.~\ref{plot:cosphi:mainfit-C} and~\ref{plot:cos2phi:mainfit-C}, for the COMPASS data, 
and in Figs.~\ref{plot:cosphi:mainfit-H} and~\ref{plot:cos2phi:mainfit-H}, for the HERMES data. 
The description of the multiplicities is practically unchanged compared to Ref.~\cite{Anselmino:2013lza}, 
therefore we do not show the plots here. 
The values of $\chi^2$ and of the parameters are listed in Tables~\ref{table:mainfit-C} and~\ref{table:mainfit-H}. 

%%%%%%%%%%%%%%%%%%%%%%%%%%%%%%%%%%%%%%%%%%%%%%%%%%%%%%%%%%%%%%%%%%%%%%%
\begin{table}[t]
\caption{Minimal $\chi^2$ and parameters, 
for a fit on COMPASS multiplicities, \cosphi and \costwophi in which 
the Cahn effect in \costwophi has been set to zero.
Parameter errors correspond to a $2\sigma$ confidence level.}
\begin{tabular}{lllll}
\hline
\hline
\rule[-5pt]{0pt}{4ex} 
~ Cuts &  ~~$ \chi^2$  & ~~~ &  ~~Parameters & \\
\hline
\rule[0pt]{0pt}{4ex}
~ $ z_h < 0.65\mathtab \mathtab$ & ~$\,(\chi^2_{pt})_{mult}\,= 3.43$  & ~ & ~$    ~~A         = 0.200\pm 0.002 $ & $  ~~M_{BM}^2  = 0.09\pm 0.45  $\\
~ $ Q^2 > 1.69\mathtab \mathtab$ & ~$(\chi^2_{pt})_{cos(\phi)}= 1.17$ & ~ & ~$    ~~B         = 0.571\pm 0.018 $ & $  ~~N_d       =-1.00\pm 1.95  $\\
~ $ 0.2  <  P_T < 0.9\mathtab$   & ~$(\chi^2_{pt})_{cos(2\phi)}=1.02$ & ~ & ~$    ~~C         = 0.031\pm 0.006 $ & $  ~~N_u       =-0.45\pm 0.26$\rule[-5pt]{0pt}{4ex}\\
\hline
\hline
\end{tabular}
\label{table:mainfit-C}
\end{table}
%%%%%%%%%%%%%%%%%%%%%%%%%%%%%%%%%%%%%%%%%%%%%%%%%%%%%%%%%%%%%%%%%%%%%%

%%%%%%%%%%%%%%%%%%%%%%%%%%%%%%%%%%%%%%%%%%%%%%%%%%%%%%%%%%%%%%%%%%%%%%
\begin{table}[t]
\caption{Minimal $\chi^2$ and parameters, 
for a fit on HERMES multiplicities, \cosphi and \costwophi in which 
the Cahn effect in \costwophi has been set to zero.
Parameter errors correspond to a $2\sigma$ confidence level.}

\begin{tabular}{lllll}
\hline
\hline
\rule[-5pt]{0pt}{4ex} 
~ Cuts &  ~~$ \chi^2$  & ~~~~ &  ~~Parameters & \\
\hline
\rule[0pt]{0pt}{4ex}
~ $ z_h < 0.69\mathtab \mathtab$ & ~$\,(\chi^2_{pt})_{mult}\,= 1.70$  & ~ & ~$    A         = 0.126\pm 0.004 $ & $  ~~M_{BM}^2  = 0.10\pm 0.20  $\\
~ $ Q^2 > 1.69\mathtab \mathtab$ & ~$(\chi^2_{pt})_{cos(\phi)}= 2.39$ & ~ & ~$    B         = 0.506\pm 0.045 $ & $  ~~N_d       =-1.00\pm 0.20  $\\
~ $ 0.2  <  P_T < 0.9\mathtab$   & ~$(\chi^2_{pt})_{cos(2\phi)}=2.13$ & ~ & ~$    C         = 0.037\pm 0.004 $ & $  ~~N_u      =-0.49\pm 0.15$\rule[-5pt]{0pt}{4ex}\\
\hline
\hline
\end{tabular}
\label{table:mainfit-H}
\end{table}
%%%%%%%%%%%%%%%%%%%%%%%%%%%%%%%%%%%%%%%%%%%%%%%%%%%%%%%%%%%%%%%%%%%%%%

The asymmetry data (especially \cosphi) drive the 
transverse momentum of quarks to 
a very small value, $\avk \sim 0.03-0.04$ GeV$^2$, which means that 
the transverse momentum of the produced hadron is largely 
due to transverse motion effects in the fragmentation process. 
The difference between positive and negative 
hadrons is found to be small (or even negligible), 
and the agreement with the data on $\langle \cos \phi \rangle$ worsens 
as $z_h$ grows and $Q^2$ decreases. 

One may wonder whether a more flexible model, including
flavor dependent Gaussian widths, could modify our results.
It would indeed be interesting to determine whether the available SIDIS data
signal any flavor dependence in the unpolarized TMDs. 
This was already considered in the analysis of multidimensional multiplicities of Ref.~\cite{Anselmino:2013lza}, where
it was noted that flavor dependence did not improve the description of the data significantly. 
Here we have performed several fits of the asymmetries allowing for 
flavor dependent parameters, but we have found that the overall picture does not improve. 
Furthermore including flavor dependence generates an over-parameterization, given the precision of present data, and consequently results in largely unconstrained
fit solutions.

From Tables~\ref{table:mainfit-C} and~\ref{table:mainfit-H} one sees that the presence 
of the Boer--Mulders function is rather marginal: $\Delta f_{d^{\uparrow}/p}$ 
is very uncertain and compatible with zero, whereas $\Delta f_{u^{\uparrow}/p}$ is 
slightly more constrained, but very small. 

The reason of this result is that 
our selection of multidimensional data on $\cos 2 \phi$, 
which cuts out a large portion of data corresponding to small $Q^2$ values, 
turns out to be compatible with a zero asymmetry.   
Notice however that in the small $Q^2$ and large $z$ region the asymmetry is instead quite sizable. 
This can be seen also by considering the previous, 
one-dimensional data, where all variables but one are integrated 
over~\cite{Kafer:2008ud,Bressan:2009eu,Giordano:2009hi}.
This means that the integrated asymmetries are mainly driven by small $Q^2$ and large $z$ events, 
which could be affected by relevant higher-twist contributions.
The importance of the $\mathcal{O}(1/Q^2)$ Cahn term, Eq.~(\ref{g-FUUcos2phi_cahn}), 
was indeed pointed out in Ref.~\cite{Barone:2009hw}, but one should not forget that 
this term is only a part of the overall twist-four contribution,  which 
is not explicitly known (at this order there are also target-mass effects, 
and the identification of $x_B$ and $z_h$ with the 
light-cone ratios $x$ and $z$ is no more valid).
%
%
%%%%%%%%%%%%%%%%%%%%%%%%%%%%%%%%%%%%%%%%%%%%%%%%%%%%%%%%%%%%%%%%%%%%%%
%%%%%%%%%%%%%%%%%%%%%%%%%%%%%%%%%%%%%%%%%%%%%%%%%%%%%%%%%%%%%%%%%%%%%%
%%%%%%%%%%%%%%%%%%%%%%%%%%%%%%%%%%%%%%%%%%%%%%%%%%%%%%%%%%%%%%%%%%%%%%
\begin{figure}[t]
    \centering
\includegraphics[width=13.5cm]{HERMESAZIM-2cos2phi-vs-PT-z-3-main-hermes-zmax0_67}
    \caption{Best fit curves for \costwophi obtained by fitting HERMES data
    on multiplicities, \cosphi and \costwophi. 
    The Cahn effect in \costwophi has been set to zero. Here we show only one bin in $z$, as an example, with $\langle z \rangle = 0.45$}
    \label{plot:cos2phi:mainfit-H}
\end{figure}
%%%%%%%%%%%%%%%%%%%%%%%%%%%%%%%%%%%%%%%%%%%%%%%%%%%%%%%%%%%%%%%%%%%%%%
%
%

By inspecting the recent release of azimuthal asymmetries by HERMES~\cite{Airapetian:2012yg} and COMPASS~\cite{Adolph:2014pwc}, 
the difference between the sizes of the negative and positive hadron production
asymmetries is evident in the one-dimensional data (see Figs. $14$--$16$ of Ref.~\cite{Airapetian:2012yg} and in 
Figs. $14$, $15$ of Ref.~\cite{Adolph:2014pwc}); 
moreover, in this case both $A^{\cos \phi}$ and $A^{\cos 2\phi}$ are consistently different from 
zero. However, these same features are not readily visible in the multidimensional data sets.
%shown %as a function of $P_T$ in Figs.~\ref{plot:compass:cosphiall} and~\ref{plot:compass:cos2phiall} and 
%as a function of $x_B$ 
%in Figs. $12$, $13$ of Ref.~\cite{Adolph:2014pwc}. 

At this stage we would like to understand whether we can assign a physical meaning to the extracted parameters.
As a matter of fact, the values of the average momentum $\avk$ extracted from our fit is quite different 
(one order of magnitude smaller) from that previously extracted in several different 
fits~\cite{Anselmino:2005nn,Signori:2013mda,Anselmino:2013lza}.
%, and they turn out to be somehow difficult 
%to interpret in terms of Gaussian widths, $\avk$ and $\avp$.
This small value is mainly driven by the \cosphi asymmetry, which 
is entirely twist-three. 
It is clear that one should make sure to evaluate this asymmetry correctly in 
order to interpret the extracted parameter as the average momentum $\avk$.

The Cahn and Boer--Mulders terms 
in $\langle \cos \phi \rangle$ can, in fact, be supplemented by other 
dynamical $1/Q$ contributions~\cite{Bacchetta:2006tn}. 
Taking quark-gluon correlations into account Eq.~(\ref{cosphiCahn}) acquires, for instance, an additional term 
containing a new distribution function, $\widetilde{f}_q$, 
%and a new fragmentation function, $\widetilde{D}_q(z,p_\perp)$, 
and becomes
\be
\left. F_{UU}^{\cos\phi} \right \vert_{\rm Cahn} = 
- 2 \sum_q e_q^2 x \int \D^2\bfk_{\perp}\,\frac{(\bfk_{\perp}\cdot \bfh)}{Q}
[f_q(x,\kt) + \widetilde{f}_q (x, \kt)]  D_q(z,p_\perp).  
\ee
%
%
% \be
% \left. F_{UU}^{\cos\phi} \right \vert_{\rm Cahn} = 
% - 2 \sum_q e_q^2 x \int \D^2\bfk_{\perp}\,\left\{ \frac{(\bfk_{\perp}\cdot \bfh)}{Q}
% [f_q(x,\kt) + \widetilde{f}_q (x, \kt)]  D_q(z,p_\perp) +  \frac{(\bfp_{\perp}\cdot \bfh)}{Q} \, f_q (x, \kt) \widetilde{D}_q(z,p_\perp)\right\}.  
% \ee
%
Notice that this new term cannot be separated from the other.
% In principle, an analogous contribution to the $\langle \cos \phi \rangle$ Cahn term would arise from the twist-three fragmentation function 
% $\widetilde{D}_q$ coupled to the twist-two distribution function $f_q (x, \kt)$, while similar corrections, proportional to the 
% $\widetilde{h}_q$ and to the fragmentation function $\widetilde{H}_q$ would   
% apply to the Boer--Mulders component~(\ref{cosphiBM}).
A similar correction, with another distribution function
$\widetilde{h}_q$,  applies to the Boer--Mulders component~(\ref{cosphiBM}), and the fragmentation functions should also be modified  
for quark-gluon correlations.

Whereas it is generally believed that 
the ``tilde'' $\mathcal{O}(1/Q)$ contributions are negligible, it is possible 
that in the kinematical regions presently explored this might not be the case. 
Possible cancellations among these terms could affect the extraction 
%This could affect our extractions 
of the intrinsic momenta parameters.

In order to estimate the impact of the dynamical higher-twist contributions to the Cahn effect, 
we can simply assume tilde functions to be proportional to the non-tilde ones.
%$\tilde f_q(x,\kt)$ to be proportional to $f_q(x,\kt)$.
Given the restricted kinematical ranges of the data, this is not a very 
limiting assumption. 
Thus the presence of $\widetilde{f}_q$ %and of $\widetilde{D}_q$ 
is effectively simulated by an extra normalization constant $N_{\rm Cahn}$ in 
front of the \cosphi Cahn term, that is: 
\[
  A^{\cos \phi} = N_{\rm Cahn} \left. A^{\cos \phi} \right \vert_{\rm Cahn} + 
 \left. A^{\cos \phi} \right \vert_{\rm BM}\,. 
\]
Since \cosphi is dominant in our fit, one sees from Eq.~(\ref{g-FUUcosphi_cahn}) that 
the effect of $N_{\rm Cahn}$ is compensated by a readjustment 
of $\avk$ (which determines not only the width of the Gaussian
distributions, but also the size of the asymmetries). Therefore, 
identical fits are obtained by allowing $N_{\rm Cahn}$ to be smaller 
than unity, and proportionally increasing $\avk$. For instance, 
setting $N_{\rm Cahn} = 0.5$, that is assuming that 
dynamical twist--3 terms reduce by 50\% the Cahn term, 
one gets $\avk = 0.06$ GeV$^2$ (twice the value obtained 
in our main fit). An even larger cancellation, that is a smaller $N_{\rm Cahn} \sim 0.1$ coefficient, 
would deliver a value of $\avk$ similar to those extracted in previous 
analyses~\cite{Anselmino:2005nn,Schweitzer:2010tt,Signori:2013mda}.

We conclude that in the present kinematics the structure and the magnitude 
of the higher--twist terms, which are not fully under control, are crucial for 
determining $\avk$. 
 
\newpage

\section{Conclusions and perspectives}

In the TMD framework, the \cosphi and \costwophi asymmetries 
are sensitive to the transverse momentum of quarks inside the target and in the 
fragmentation process. In the Gaussian model of quark distributions, 
the widths $\avk$ and $\avp$ also determine  the size of the 
asymmetries. Adopting an $\mathcal{O} (1/Q)$ 
scheme, which attributes \cosphi both to Cahn and Boer--Mulders effects at order 
$1/Q$, and \costwophi to the Boer--Mulders effect at leading twist, and ignoring 
twist-3 dynamical contributions (arising from quark--gluon correlations), our analysis shows 
that the recent COMPASS and HERMES multidimensional data can be reproduced 
by a very small value of $\avk$, namely 0.03-0.04 GeV$^2$. Within this picture, this means that 
most of the transverse momentum of the outgoing hadron is due to 
the fragmentation, which must be described by a function with a $z$-dependent 
width.  This result, mainly driven by \cosphi, could be modified 
by the presence of further twist--3 terms, 
which might not be negligible due to the relevance  of the 
small-$Q^2$ region in the present measurements. 

%In principle, both azimuthal asymmetries can be generated perturbatively by gluon radiation at 
%order $\alpha_s$~\cite{Georgi:1977tv,Mendez:1978zx,Konig:1982uk,Chay:1991nh}, but we have 
%explcitely checked that this contribution is irrelevant in the kinematical regime of the SIDIS 
%experiments we have considered here.

A somehow disappointing output of our fits is the indeterminacy 
on the extraction of the Boer--Mulders function, which seems to
play a minor role in the asymmetries. 
This is seen in particular from \costwophi, which is entirely determined by the 
Boer--Mulders contribution but appears to be, within large errors,  
compatible with zero.   

On the other hand, the integrated \costwophi data~\cite{Airapetian:2012yg}
show a non vanishing asymmetry, especially when plotted
against $z$. The asymmetry is slightly negative for $\pi^+$ and positive
for $\pi^-$,
as expected from the Boer--Mulders effect \cite{Barone:2008tn}.
Also the integrated data on \cosphi show a different asymmetry
for $\pi^+$ and $\pi^-$:
this indicates a flavor 
dependence which can only be achieved with a non-zero Boer--Mulders effect since, 
within a flavor--independent Gaussian model with factorized $x$ and $\kt$ dependences, 
the Cahn effect is flavor blind and can only generate 
identical contributions for positively or negatively charged pions.
However, the signs of the $u$ and $d$ Boer--Mulders functions required for a successful description 
of \costwophi appear to be incompatible with those required to generate the appropriate difference between  
$\pi^+$ and $\pi^-$ in the \cosphi azimuthal moment. 
As we mentioned, a more refined model with flavor dependent Gaussian widths is not helpful, 
given the precision of the current experimental data.

One should not forget about the existence of other higher-twist effects that 
could combine with the Boer--Mulders term and alter the simple picture considered here. 
In order to disentangle these 
contributions, it might be useful to integrate the asymmetry data 
on restricted kinematical ranges, so as to avoid the low-$Q^2$ region 
and meet the requirements of TMD factorization. 
Analyzing properly integrated data could help to clarify 
the origin of azimuthal asymmetries 
and possibly to get more information on the Boer--Mulders function. 
Work along these lines is in progress.

It would also be interesting to investigate how SIDIS azimuthal modulations can be affected by gluon radiations. 
Following Ref.~\cite{Chay:1991nh} one can actually compute the perturbative corrections 
originating from gluon radiation at order $\alpha_s$ for the numerator and denominator of the azimuthal asymmetries.
Indeed, in the limit of small $q_T$ (where our analysis applies) they are affected by strong divergences, generated by soft 
and collinear gluon radiation. One might expect that these divergences cancel out in the ratio when building the 
azimuthal asymmetry, but this only happens when the divergences appearing in the numerator and in the denominator 
are of the same origin and have the same structure.
In fact, in a more recent paper~\cite{Berger:2007jw}, it is explicitly shown that, 
for Drell-Yan scattering processes, the \costwophi azimuthal modulation of the cross section shares with the azimuthal 
independent (integrated) term the same $q_T$ logarithmic behaviour of the asymptotic cross section, proportional 
to $(Q^2/q_T^2)\ln(Q^2/q_T^2)$. 
In this case the same resummation techniques which are known to work for the integrated cross section (which appear at denominator) 
can be applied to the azimuthal modulation appearing at numerator. However, they point out that this does not happen for the \cosphi 
azimuthal modulation, which does not exhibit the usual diverging logarithmic term, but is simply proportional to $(Q^2/q_T^2)$.
In this case, the usual resummation scheme techniques cannot be applied and new strategies needs to be devised.
To the best of our knowledge, this has not been explicitly studied for SIDIS processes, where applying usual resummation schemes 
and conventional matching recipes is much more problematic, even in the simplest case of integrated, unpolarized cross sections~\cite{Boglione:2014oea}.

\begin{acknowledgments}

We are grateful to Mauro Anselmino for his support and for the endless conversations on the subject, 
and to Anna Martin and Franco Bradamante for useful discussions on COMPASS data. 
This work was partially supported by the European Research Council under the FP7
``Capacities - Research Infrastructure'' program (HadronPhysics3, Grant Agreement 283286).
M.B., J.O.G.H. and S.M. are also supported by the ``Progetto di Ricerca Ateneo/CS'' (TO-Call3-2012-0103).
V.B. gratefully acknowledges the hospitality 
of the Physics Department of the Universit{\`a} di Trieste, 
where part of this work was done during a sabbatical leave 
in 2014. 

\end{acknowledgments}
%%%%%%%%%%%%%%%%%%%%%%%%%%%%%%%%%%%%%%%%%%%%%%%%%%%%%%%%%%%%%%%%%%%%%%%%%%%%%%%

\bibliography{azimut_biblio}{}

\begin{thebibliography}{10}

\bibitem{Georgi:1977tv}
H.~Georgi and H.~D. Politzer,
\newblock Phys. Rev. Lett. {\bf 40}, 3 (1978).
%%CITATION = PRLTA,40,3;%%

\bibitem{Mendez:1978zx}
A.~Mendez,
\newblock Nucl. Phys. {\bf B145}, 199 (1978).
%%CITATION = NUPHA,B145,199;%%

\bibitem{Cahn:1978se}
R.~N. Cahn,
\newblock Phys. Lett. {\bf B78}, 269 (1978).
%%CITATION = PHLTA,B78,269;%%

\bibitem{Cahn:1989yf}
R.~N. Cahn,
\newblock Phys. Rev. {\bf D40}, 3107 (1989).
%%CITATION = PHRVA,D40,3107;%%

\bibitem{Konig:1982uk}
A.~Konig and P.~Kroll,
\newblock Z. Phys. {\bf C16}, 89 (1982).
%%CITATION = ZEPYA,C16,89;%%

\bibitem{Chay:1991nh}
J.-g. Chay, S.~D. Ellis, and W.~J. Stirling,
\newblock Phys. Rev. {\bf D45}, 46 (1992).
%%CITATION = PHRVA,D45,46;%%

\bibitem{Barone:2010zz}
V.~Barone, F.~Bradamante, and A.~Martin,
\newblock Prog.Part.Nucl.Phys. {\bf 65}, 267 (2010), arXiv:1011.0909.
%%CITATION = ARXIV:1011.0909;%%

\bibitem{Ji:2004wu}
X.-d. Ji, J.-p. Ma, and F.~Yuan,
\newblock Phys. Rev. {\bf D71}, 034005 (2005), arXiv:hep-ph/0404183.
%%CITATION = HEP-PH/0404183;%%

\bibitem{Aubert:1983cz}
European Muon, J.~J. Aubert {\em et~al.},
\newblock Phys. Lett. {\bf B130}, 118 (1983).
%%CITATION = PHLTA,B130,118;%%

\bibitem{Arneodo:1986cf}
European Muon, M.~Arneodo {\em et~al.},
\newblock Z. Phys. {\bf C34}, 277 (1987).
%%CITATION = ZEPYA,C34,277;%%

\bibitem{Breitweg:2000qh}
ZEUS, J.~Breitweg {\em et~al.},
\newblock Phys. Lett. {\bf B481}, 199 (2000), arXiv:hep-ex/0003017.
%%CITATION = HEP-EX/0003017;%%

\bibitem{Boer:1997nt}
D.~Boer and P.~J. Mulders,
\newblock Phys. Rev. {\bf D57}, 5780 (1998), arXiv:hep-ph/9711485.
%%CITATION = HEP-PH/9711485;%%

\bibitem{Kafer:2008ud}
COMPASS Collaboration, W.~Kafer,
\newblock (2008), arXiv:0808.0114.
%%CITATION = ARXIV:0808.0114;%%

\bibitem{Bressan:2009eu}
COMPASS Collaboration, A.~Bressan,
\newblock (2009), arXiv:0907.5511.
%%CITATION = ARXIV:0907.5511;%%

\bibitem{Giordano:2009hi}
HERMES Collaboration, F.~Giordano and R.~Lamb,
\newblock AIP Conf.Proc. {\bf 1149}, 423 (2009), arXiv:0901.2438.
%%CITATION = ARXIV:0901.2438;%%

\bibitem{Barone:2009hw}
V.~Barone, S.~Melis, and A.~Prokudin,
\newblock Phys.Rev. {\bf D81}, 114026 (2010), arXiv:0912.5194.
%%CITATION = ARXIV:0912.5194;%%

\bibitem{Barone:2008tn}
V.~Barone, A.~Prokudin, and B.-Q. Ma,
\newblock Phys.Rev. {\bf D78}, 045022 (2008), arXiv:0804.3024.
%%CITATION = ARXIV:0804.3024;%%

\bibitem{Airapetian:2012ki}
HERMES Collaboration, A.~Airapetian {\em et~al.},
\newblock Phys.Rev. {\bf D87}, 074029 (2013), arXiv:1212.5407.
%%CITATION = ARXIV:1212.5407;%%

\bibitem{Adolph:2013stb}
COMPASS, C.~Adolph {\em et~al.},
\newblock Eur.Phys.J. {\bf C73}, 2531 (2013), arXiv:1305.7317.
%%CITATION = ARXIV:1305.7317;%%

\bibitem{Airapetian:2012yg}
HERMES Collaboration, A.~Airapetian {\em et~al.},
\newblock Phys.Rev. {\bf D87}, 012010 (2013), arXiv:1204.4161.
%%CITATION = ARXIV:1204.4161;%%

\bibitem{Adolph:2014pwc}
COMPASS Collaboration, C.~Adolph {\em et~al.},
\newblock Nucl.Phys. {\bf B886}, 1046 (2014), arXiv:1401.6284.
%%CITATION = ARXIV:1401.6284;%%

\bibitem{Schweitzer:2010tt}
P.~Schweitzer, T.~Teckentrup, and A.~Metz,
\newblock Phys.Rev. {\bf D81}, 094019 (2010), arXiv:1003.2190.
%%CITATION = ARXIV:1003.2190;%%

\bibitem{Musch:2007ya}
LHPC Collaboration, B.~U. Musch {\em et~al.},
\newblock PoS {\bf LAT2007}, 155 (2007), arXiv:0710.4423.
%%CITATION = ARXIV:0710.4423;%%

\bibitem{Pumplin:2002vw}
J.~Pumplin {\em et~al.},
\newblock JHEP {\bf 0207}, 012 (2002), arXiv:hep-ph/0201195.
%%CITATION = HEP-PH/0201195;%%

\bibitem{deFlorian:2007aj}
D.~de~Florian, R.~Sassot, and M.~Stratmann,
\newblock Phys.Rev. {\bf D75}, 114010 (2007), arXiv:hep-ph/0703242.
%%CITATION = HEP-PH/0703242;%%

\bibitem{Anselmino:2013lza}
M.~Anselmino, M.~Boglione, J.~Gonzalez~H., S.~Melis, and A.~Prokudin,
\newblock JHEP {\bf 1404}, 005 (2014), arXiv:1312.6261.
%%CITATION = ARXIV:1312.6261;%%

\bibitem{Anselmino:2013vqa}
M.~Anselmino {\em et~al.},
\newblock Phys.Rev. {\bf D87}, 094019 (2013), arXiv:1303.3822.
%%CITATION = ARXIV:1303.3822;%%

\bibitem{Stolarski:2015}
COMPASS Collaboration, M.~Stolarski,
\newblock Contribution to the 21st International Symposium on Spin Physics
  (SPIN 2014), October 20-24, 2014, Beijing, China  (2015).

\bibitem{Anderle:2012rq}
D.~P. Anderle, F.~Ringer, and W.~Vogelsang,
\newblock Phys.Rev. {\bf D87}, 034014 (2013), arXiv:1212.2099.
%%CITATION = ARXIV:1212.2099;%%

\bibitem{Anselmino:2005nn}
M.~Anselmino {\em et~al.},
\newblock Phys.Rev. {\bf D71}, 074006 (2005), arXiv:hep-ph/0501196.
%%CITATION = HEP-PH/0501196;%%

\bibitem{Signori:2013mda}
A.~Signori, A.~Bacchetta, M.~Radici, and G.~Schnell,
\newblock JHEP {\bf 1311}, 194 (2013), arXiv:1309.3507.
%%CITATION = ARXIV:1309.3507;%%

\bibitem{Bacchetta:2006tn}
A.~Bacchetta {\em et~al.},
\newblock JHEP {\bf 0702}, 093 (2007), arXiv:hep-ph/0611265.
%%CITATION = HEP-PH/0611265;%%

\bibitem{Berger:2007jw}
E.~L. Berger, J.-W. Qiu, and R.~A. Rodriguez-Pedraza,
\newblock Phys.Rev. {\bf D76}, 074006 (2007), arXiv:0708.0578.
%%CITATION = ARXIV:0708.0578;%%

\bibitem{Boglione:2014oea}
M.~Boglione, J.~O.~G. Hernandez, S.~Melis, and A.~Prokudin,
\newblock JHEP {\bf 1502}, 095 (2015), arXiv:1412.1383.
%%CITATION = ARXIV:1412.1383;%%

\end{thebibliography}
\bibliographystyle{h-physrev5}

\end{document}